\documentclass[12pt]{article}
\usepackage{amsfonts,latexsym,amsmath}
\title{Symmetries and Lagrangian time-discretizations of Euler equations}
\author{Alexei V. Penskoi\thanks{Centre de recherches math\'ematiques, 
Universit\'e de Montr\'eal,
C.~P.~6128, succ. Centre-ville, Montr\'eal, 
Qu\'ebec, H3C 3J7, Canada {\tt e-mail: penskoi@crm.umontreal.ca}}}
\date{}
\newcommand{\diff}{\mbox{Diff}_+(S^1)}
\newtheorem{theorem}{Theorem}

\begin{document}
\maketitle

\abstract{In the late 80s - early 90s J. Moser and A. P. Veselov considered 
Lagrangian discrete systems on Lie groups with additional symmetry conditions 
imposed on Lagrangians. They observed that such systems are often integrable 
time-discretizations of integrable Euler equations on these Lie groups. In 
recent papers we studied Lagrangian discrete systems with additional symmetry 
requirements on certain infinite-dimensional Lie groups. We will discuss some 
interesting properties of these systems. }

\noindent 2000 Mathematical Subject Classification 34K99, 22E65, 70H99

\noindent {\bf Keywords:} Euler equation, Discrete Lagrangian system, 
Virasoro group, Camassa-Holm equation, Korteweg-de~Vries equation,
Hunter-Saxton equation.

\section{Introduction}

The goal of this talk is to review some recent results in discrete Lagrangian
systems~\cite{P1,P2,P3,PV}.

Let $M$ be a manifold and $L$ be a function on $M\times M.$
The {\it discrete Lagrangian system} with the Lagrangian $L$
is the system of difference equations
$$ \delta S =0,$$
which describes the
stationary points of the functional
$S=S(X),$ defined on the space of sequences $X=(x_k),x_k \in M,
k\in{\mathbb Z}$ by
a formal sum
$$
S=\sum\limits_{k\in{\mathbb Z}}L(x_k,x_{k+1}).
$$
Sometimes this system has a continuous limit; in that case it is called
a discrete version of the corresponding system of differential equations.

For integrable equations it is natural to ask for discretizations
which are also integrable. The theory of integrable Lagrangian
discretizations of classical integrable systems
was initiated by Moser and Veselov~\cite{V1, MV1}.

Let us assume that $M$ is a Lie group $G.$ It was observed by Veselov
and Moser~\cite{V1,MV1,MV2,V2} that symmetric
$$
L(x,y)=L(y,x)
$$
and right-invariant (or left-invariant)
$$
\forall g\in G\quad L(xg,yg)=L(x,y)
$$
Lagrangians often correspond to integrable systems which are
discretizations of the Euler equations corresponding to some right-invariant
(or left-invariant) metrics on $G.$

In particular, it was shown that the discrete Lagrangian system
on the orthogonal group $O(n)$ with the Lagrangian
$L(X,Y) =tr( XJY^{T}),$ where $X,Y\in O(n)$ and $J$ is a
symmetric positive-defined matrix, can be considered
as an integrable discrete version of the Euler-Arnold top \cite{A}.

The first attempt to generalize this approach to the infinite-dimensional
situation was done in \cite{MV2}, where the case of the group of
area-preserving
plane diffeomorphisms $\mbox{SDiff}({\mathbb R}^2)$ was considered.

This paper was followed by recent papers~\cite{P1,P2,P3,PV} where
discrete Lagrangian systems on the Virasoro group have been discussed.
The interest in
the Virasoro group was motivated by the important observation due to
Khesin and Ovsienko~\cite{KO} that the Korteweg-de~Vries
equation can be interpreted as an Euler equation on the Virasoro algebra.
This result was later generalized by Misio\l{}ek~\cite{M} and
Khesin~\cite{KM}. They remarked that
the Euler equation corresponding to the right-invariant
$H^1_{\alpha,\beta}$-metric on the Virasoro group has the form
$$
\alpha(v_t+3vv_x)-\beta(v_{xxt}+2v_xv_{xx}+ vv_{xxx})-bv_{xxx}=0.
$$
This is a three-parameter family of integrable equation including
the Camassa-Holm, the Hunter-Saxton and the Korteweg-de~Vries equations.

The plan of the talk is following. We shall start by recalling the basic facts
about the Euler equation in Section~\ref{euler} and the Virasoro 
group in Section~\ref{virga}. Then we will expose in Section~\ref{okm}
results of Ovsienko, Khesin and Misio\l{}ek~\cite{KO,M,KM}
concerning Euler equations on the Virasoro group. Section~\ref{discr}
is devoted to discrete Lagrangian systems, and 
the final Section~\ref{trudy}
contains results on discrete Lagrangian systems on the Virasoro group.

\section{The Euler equation}\label{euler}

Let $G$ be a Lie group, $\mathfrak{g}$ its Lie algebra. A right-invariant
metric on $G$ is completely defined by its restriction to $\mathfrak{g}$
$$
(\,,):\mathfrak{g}\times\mathfrak{g}\rightarrow\mathbb{R}.
$$
This scalar product on $\mathfrak{g}$ defines a linear map 
$A:\mathfrak{g}\longrightarrow\mathfrak{g}^*$ such that
$$
(\xi,\eta)=\langle A\xi,\eta\rangle,
$$
where $\xi,\eta\in\mathfrak{g}$ and $\langle\,,\rangle$ denotes the natural 
pairing between $\mathfrak{g}$ and $\mathfrak{g}^*.$ The operator $A$ is
called the inertia operator.

To describe a geodesic $g(t)$ on $G$ corresponding to $(\,,),$
we transport its velocity vector to the identity by the right translation
$$
v(t)=R_{g^{-1}(t)*}\frac{d}{dt}g(t).
$$
Since $v(t)$ is an element of $\mathfrak{g},$ we can consider $m=Av$ which
is an element of the dual space $\mathfrak{g}^*.$ Then $m$ satisfies
the Euler equation given by the following explicit formula:
\begin{equation}\label{eulereq}
\frac{dm}{dt}=-ad^*_{A^{-1}m}m.
\end{equation}
This is a standard result which can be found in~\cite{A}. If we start with
a left-invariant metric, the sign in~(\ref{eulereq}) is reversed.
Since $A$ is non-degenerate, one can rewrite the Euler equation in terms of
$v$ on the Lie algebra $\mathfrak{g}.$

{\bf Example 1.}~\cite{A} Let us consider $SO(N).$ We identify the dual space
$\mathfrak{so}(N)^*$ with the algebra $\mathfrak{so}(N)$ using the
standard product $\mbox{tr}XY^T$ in the space of $N\times N$ matrices.
Let $J$ be a symmetric positive-definite matrix. 
The inertia operator defined as
$$A\Omega=J\Omega+\Omega J,\quad \Omega\in\mathfrak{so}(N)$$
gives the Euler equation on $SO(N).$ It can be written in a more familiar 
form as a system
$$
\left\{
\begin{array}{rcl}
\dot{M}&=&[M,\Omega],\\
M&=&J\Omega+\Omega J.
\end{array}
\right.
$$
This is the Euler-Arnold equation for the $N$-dimensional rigid body.

{\bf Example 2.}~\cite{A} 
Let us consider the group $\mbox{SDiff}(\mathbb{R}^2)$ 
of volume-preserving diffeomorphisms of $\mathbb{R}^2.$ The corresponding
Lie algebra is the algebra of all vector fields with divergence $0$
on $\mathbb{R}^2.$ Let us consider the scalar product of two such
vector fields defined as
$$
(v_1,v_2)=\iint_{\mathbb{R}^2}v_1\cdot v_2\,dx_1dx_2,
$$
where $\cdot$ denotes the standard scalar product of vectors in
$\mathbb{R}^2.$ We can find the Euler equation corresponding to this scalar
product. We will write it on the Lie algebra in the following form:
$$
\left\{
\begin{array}{rcl}
\frac{\partial v}{\partial t}+v\nabla v&=&-\mbox{grad}\,p,\\
\quad \mbox{div}\,v&=&0.
\end{array}
\right.
$$
This is the two-dimensional Euler equation for the ideal fluid. Here $p$
is a pressure, it is some unknown function which plays the role of a 
Lagrangian multiplier defined by the constraint $\mbox{div}\,v=0.$

\section{The Virasoro group and the Virasoro algebra}\label{virga}

Let  $\diff$ be the group of diffeomorphisms of $S^1$ preserving 
the orientation. We shall represent an element of $\diff$ as a diffeomorphism
$f:{\mathbb R}\rightarrow{\mathbb R}$ such that 
\begin{enumerate}
\item $f\in C^\infty({\mathbb R}),$
\item $f'(x)>0,$
\item $f(x+2\pi)=f(x)+2\pi.$
\end{enumerate}
Such a representation is not unique. Indeed, the functions 
$f+2\pi k,k\in\mathbb{Z}$ 
represent one element of $\diff.$

There exists a non-trivial central extension of
$\diff$ which is unique up to an isomorphism

This extension is called the Virasoro group (or the Bott-Virasoro group) 
and is denoted by
$\mbox{Vir}.$ Elements of $\mbox{Vir}$ are pairs $(f,F)$, 
where $f\in\diff,$ $F\in{\mathbb R}.$
The product of two elements is defined 
with the help of the Bott cocycle
as
$$
(f,F)\circ(g,G)=(f\circ g,F+G+\int\limits_{0}^{2\pi}\log(f\circ g)'\,d\log g').
$$
The unit element of $\mbox{Vir}$ is $(id,0).$ The inverse element of $(f,F)$ is
$(f^{-1},-F).$

The Virasoro algebra $\mbox{vir}$
is a Lie algebra corresponding to the Virasoro group. It is the 
central extension of the algebra $vect(S^1)$ of vector fields on 
the circle $S^1$
$$
\mbox{vir}=vect(S^1)\oplus{\mathbb R}.
$$
We represent an element of $vect(S^1)$ as $v(x)\partial_x,$ where
$v$ is a $2\pi$-periodic function. Thus an element of the Virasoro
algebra is a pair $(v(x)\partial_x,a).$ The algebra commutator in $\mbox{vir}$
is defined 
with the help of the Gelfand-Fuchs cocycle as
$$
[(v(x)\partial_x,a],[w(x)\partial_x,b])=%
((-vw_x+v_xw)(x)\partial_x,\int\limits_0^{2\pi}v_{xxx}w\,dx).
$$

\section{The Euler equation for the $H^1_{\alpha,\beta}$-metric on the 
Virasoro group}\label{okm}

In 1987 Khesin and Ovsienko remarked~\cite{KO} that the Korteweg-de~Vries
equation can be interpreted as an Euler equation on the Virasoro group.
This result was later generalized by Misio\l{}ek~\cite{M} and
Khesin~\cite{KM} in the following way.

Let $\alpha$ and $\beta$ be two non-negative
real numbers such that $\alpha^2+\beta^2\ne 0.$
Let us define the $H^1_{\alpha,\beta}$-metric on the 
Virasoro algebra by the formula
$$
((v(x)\partial_x,a),(w(x)\partial_x,b))_{H^1_{\alpha,\beta}}=%
\int_0^{2\pi}(\alpha v(x)w(x)+\beta v_x(x)w_x(x))\,dx+ab.
$$
\begin{theorem} 
The Euler equation corresponding to the right-invariant
$H^1_{\alpha,\beta}$-metric on the Virasoro group has the form
\begin{equation}
\alpha(v_t+3vv_x)-\beta(v_{xxt}+2v_xv_{xx}+ vv_{xxx})-bv_{xxx}=0,\label{CH1}
\end{equation}
$$
b_t=0.
$$
\end{theorem}

Since $b$ is a constant, we can consider~(\ref{CH1}) as an equation for
$v$ depending on three constants $\alpha,\beta,b.$
We obtain a three-parametric family of 
integrable equations which
we call the Camassa-Holm family. The reason for this name is the following.
We have a freedom of multiplication of the equation
by a non-zero constant, the two-dimensional scaling symmetry group
$v\rightarrow\lambda v, t\rightarrow\mu t, x\rightarrow\lambda\mu x$
and the Galilean group $v\rightarrow v + c, x\rightarrow x + dt, 
t\rightarrow t.$
Modulo these symmetries we have just one generic orbit, containing
the equation with
$\alpha = 1, \beta = 1, b =0:$
$$
v_t-v_{xxt}+3vv_x-2v_x v_{xx}-vv_{xxx}=0,
$$
which is one of the canonical forms of the Camassa-Holm
shallow-water equation~\cite{CH}
$$
v_t+2\kappa v_x+\gamma v_{xxx}-v_{xxt}+3vv_x-2v_x v_{xx}-vv_{xxx}=0.
$$

We have also four degenerate orbits.
When $\alpha\neq 0, \beta = 0, b\neq 0$
the equation (\ref{CH1}) is equivalent to the KdV equation:
$$v_t+3vv_x+v_{xxx}=0.$$
Further degeneration $\alpha\neq 0,\beta=0,b=0$ leads to the
dispersionless KdV equation (sometimes also called the Hopf equation):
$$v_t+3vv_x=0.$$
When $\alpha=0,\beta\neq 0$ we have the Hunter-Saxton
equation~\cite{HS}
$$v_{xxt}+2v_xv_{xx}+vv_{xxx}=0.$$
Finally if both $\alpha$ and $\beta$ are zero (but $b$ is not) we simply have
$$v_{xxx}=0.$$

\section{Discrete Lagrangian systems}\label{discr}

The definition of a discrete Lagrangian system is already done in
Introduction. Also, we have already mentioned in Introduction the
observation by Veselov and Moser that symmetric and right-invariant
(or left-invariant) Lagrangians often correspond to integrable systems 
which are discretizations of the Euler equations corresponding to some 
right-invariant
(or left-invariant) metrics on $G.$ Let us consider some examples.

{\bf Example 3.}~\cite{MV1} Let us consider the group $SO(N).$ Let $J$ 
be a symmetric positive-definite matrix. The Lagrangian
$$
L(X,Y)=\mbox{tr}(XJY^T)
$$
is left-invariant and symmetric. Let us introduce the discrete
angular velocity $\omega_k=X^T_kX_{k-1}$ and the discrete angular
momentum $M_k=\omega^T_kJ-J\omega_k.$ The discrete Euler-Lagrange equation
$\frac{\delta S}{\delta X_k}=0$ can be written in the following form
(discrete Euler-Arnold equation).
$$
\left\{
\begin{array}{rcl}
M_{k+1}&=&\omega_kM_k\omega^{-1}_k,\\
M_k&=&\omega^T_kJ-J\omega_k.
\end{array}
\right.
$$
This equation is integrable and its continuous limit is the Euler-Arnold
equation of $N$-dimensional rigid body considered in the Example 1.

{\bf Example 4.}~\cite{MV2,FV} Let us consider the group
$\mbox{SDiff}(\mathbb{R}^2).$ The Lagrangian
$$
L(f,g)=\iint_{\mathbb{R}^2}\mbox{tr}(J(f)J(g)^{-1})\,dx_1dx_2,
$$
where $f,g\in\mbox{SDiff}(\mathbb{R}^2)$ and $J$ is the Jacobian, is
right-invariant and symmetric. Let $\varphi=f_k\circ f^{-1}_{k-1},$
$\psi=f_{k+1}\circ f^{-1}_k$ and $\chi=\varphi^{-1}.$
The discrete Euler-Lagrange equation
$\frac{\delta S}{\delta f_k}=0$ can be written in the following form.
$$
\left\{
\begin{array}{rcl}
\psi_1&=&\tau x_1+a_1-\chi_1\\
\psi_2&=&\tau x_2+a_2-\chi_2\\
\mbox{tr}J(\varphi)&=&\tau.
\end{array}
\right.
$$
In these equations $\tau,a_1,a_2$ are constants, $\psi_i,i=1,2$
are components of $\psi=(\psi_1,\psi_2):\mathbb{R}^2\rightarrow\mathbb{R}^2,$
and $\chi_i,i=1,2$ are defined in the analogous way. 
This discrete system is integrable. If $\tau=2$ one can find the continuous 
limit
$$
\left\{
\begin{array}{rcl}
v_t+v\nabla v&=&0,\\
\mbox{div}\,v&=&0.
\end{array}
\right.
$$
This is an equation of the isobaric flow, a particular case of the 
two-dimensional Euler equation for the ideal fluid when the pressure is equal 
to zero.

\section{Discrete Lagrangian systems on the Virasoro group}\label{trudy}

In this section we present some results of papers~\cite{P1,P2,P3,PV}. Let us
consider discrete Lagrangian systems on the Virasoro group. It was explained
in Section~\ref{discr} that it is natural to consider right-invariant symmetric
Lagrangians. 

We shall use the following simple observation to construct Lagrangians.
Let us consider a Lagrangian $L(x,y)$ on a Lie group $G.$
Let $H(x)=L(x,e),$ where $e$ is the identity element of $G.$
A right-invariant Lagrangian $L(x,y)$ is completely determined by $H$.
Indeed, $L(x,y)=L(xy^{-1},e)=H(xy^{-1}).$ 

Let us now consider a right-invariant symmetric Lagrangian.
It is easy to see that symmetric Lagrangian corresponds to inverse-invariant
function $H:$
$$
H(x^{-1})=H(x).
$$

Let us start by considering the discrete Lagrangian systems on $\mbox{Vir}$ 
corresponding to the functions $H$ of the following form~\cite{PV}
\begin{equation}\label{finalH}
H((f,F))=F^2+\int\limits_0^{2\pi}V(f(x)-x,f'(x))\,dx,
\end{equation}
where $f$ is a diffeomorphism, $F\in\mathbb{R},$ and $V(x_1,x_2)$ is an
arbitrary, $2\pi$-periodic in $x_1$
function of two variables, which satisfies the condition:
$$
V_1(0,1) =0.
$$

Let us explain a motivation for such a choice of $H.$ 
We are interested in Lagrangian time-discretizations of the Camassa-Holm 
family. This family corresponds to $H^1_{\alpha,\beta}$-metrics, and
these metrics depend on elements of the algebra and their first derivatives.
So it is natural to consider a function $H$ of the form
\begin{equation}\label{functionH}
H((f,F))=F^2+\int\limits_0^{2\pi}U(f(x),f'(x),x)\,dx,
\end{equation}
where $U(x_1,x_2,x_3)$ is an arbitrary function 
that is $2\pi$-periodic with respect to
the first argument.
Periodicity of $V$ or $U$ is related to the fact that $f(x)$ and $f(x)+2\pi$
represent the same diffeomorphism of $S^1.$
However, it turns out that the continuous limit of the system corresponding 
to~(\ref{functionH}) is, in general, some quite boring ODE having
nothing in common with Euler equations.
To obtain something interesting, it is necessary to impose the additional
conditions that $U$ has the form $U(x_1,x_2,x_3)=V(x_1-x_3,x_2)$ and 
$V_1(0,1) =0.$ In this case the continuous limit is
some equation from the Camassa-Holm family.

The form~(\ref{finalH}) is very general since
the difference $f(x)-x,$ which is a $2\pi$-periodic function, is
as natural as $f(x)$ itself. The only property which might look artificial
is the last condition on the partial derivative of $V$.
See~\cite{PV} for a detailed discussion.

In general, the Lagrangian corresponding to~(\ref{finalH}) is not symmetric. 
Hence, it is not expected to give us an integrable system. Nevertheless, 
it is interesting to look at such Lagrangians because of the following 
interesting phenomenon. In general, if we consider a Lie group $G,$ 
a continuous limit of a right-invariant discrete Lagrangian system is a
geodesic flow corresponding to some right-invariant metric.  
This geodesic flow is, in general, not integrable.
Let us for example look at the case of 
$SO(N).$ It is known that for $N>3$ a geodesic flow of 
a general right-invariant
metric is non-integrable (see e.g.~\cite{V3}). 
Hence, the result on the Virasoro group turns out to be surprising:
in spite of the fact that the class of discrete systems we 
consider~(\ref{finalH}) is quite general, in the continuous limit
we have the family of integrable equations.
There are integrable cases
of the Euler equations on $SO(N)$ (for example the Manakov metrics
\cite{Manakov}) but no analogues of the result for the Virasoro group 
are known for them.

The heuristic explanation of this phenomenon is that the equations
of the Camassa-Holm family can be considered as nonlinear analogues of the 
harmonic oscillators
on the Virasoro group: in the first approximation all Hamiltonian
systems near equilibriums behave like harmonic oscillators. Thus, in some sense
this demonstrates a universal nature of the Camassa-Holm family of equations.
See Discussion in~\cite{PV} for more details.

Let us now consider the question of the
integrability of discrete Lagrangian systems
on the Virasoro group. Unfortunately we do not know examples of such systems, 
but we have good candidates for integrability. A good candidate to be
an integrable discretization of the KdV equation was considered in~\cite{P2},
but it is the case of the discretization of the Hunter-Saxton 
equation~\cite{P3} which is most promising. Let us consider this case.

The Hunter-Saxton equation corresponds to the Euler equation for 
the $H^1_{0,1}$-metric. This metric depends only on 
first derivatives of algebra 
elements. Hence, it is natural to look at such functions 
$H$~(\ref{finalH}) that $V$ depends only on $f'.$
As it was explained before, it is natural to study right-invariant symmetric
Lagrangians when looking for integrable systems, and symmetric Lagrangians
correspond to inverse-invariant functions $H.$

It can be easily verified that functions $H$ of the form 
\begin{equation}\label{HforHS}
H((f,F))=F^2+\int\limits_0^{2\pi}V(f'(x))\,dx
\end{equation}
are inverse-invariant when the function $V$ satisfies the condition
\begin{equation}\label{condU}
xV\left(\frac{1}{x}\right)=V(x).
\end{equation}
The simplest function $V$ satisfying the property~(\ref{condU}) is
$V(x)=\sqrt{x}.$ Let us consider this case.

We have $V(x)=\sqrt{x}.$ The function $V$ defines $H$ as 
described in~(\ref{HforHS}).
The function $H$ defines a Lagrangian $L$ as described above:
$$
L((f_k,F_k),(f_{k+1},F_{k+1}))=H((f_k,F_k)\circ(f_{k+1},F_{k+1})^{-1}).
$$
Hence we are considering a functional
$$
S=\sum\limits_{k\in{\mathbb Z}}L((f_k,F_k),(f_{k+1},F_{k+1})),
$$
where $\{(f_k,F_k)\}$ is a sequence of points on $\mbox{Vir}.$
We can find the Euler-Lagrange equations 
$\frac{\delta S}{\delta (f_k,F_k)}=0.$ They are~\cite{P3}:
\begin{equation}\label{EL1}
-\Omega_k+\Omega_{k+1}=0,
\end{equation}
$$
\left[-2\Omega_k(\log((\omega_k)'))'-%
\frac{1}{2}\sqrt{(\omega_k)'}+\right.
$$
\begin{equation}\label{EL2}
\left.+2\Omega_{k+1}(\log((\omega^{-1}_{k+1})'))'-%
\frac{1}{2}\sqrt{(\omega^{-1}_{k+1})'}\right]'=0,
\end{equation}
where $(\omega_k,\Omega_k)$ and $(\omega_{k+1},\Omega_{k+1})$  
are discrete analogues of angular velocities,
$$
(\omega_l,\Omega_l)=(f_{l-1},F_{l-1})\circ(f_l,F_l)^{-1},\quad l\in\mathbb{Z}.
$$
The equation~(\ref{EL1}) is just saying that $\Omega_k$ is an integral,
$\Omega_k=\Omega.$
As for~(\ref{EL2}), we can integrate this equation once
and put
$$
\Phi=\frac{1}{\sqrt{(\omega_k)'}},\quad%
\Psi=\frac{1}{\sqrt{(\omega^{-1}_{k+1})'}}.
$$
We obtain the equation
$$
8\Omega\left(-\frac{\Phi'}{\Phi}+\frac{\Psi'}{\Psi}\right)+%
\frac{1}{\Phi}+\frac{1}{\Psi}+C=0,
$$
where $C$ is a constant of integration.
This equation is equivalent to the equation
$$
\Psi'+\Psi\left(\frac{C}{8\Omega}+\frac{1}{8\Omega\Phi}-%
\frac{\Phi'}{\Phi}\right)+\frac{1}{8\Omega}=0.
$$
This is a linear first-order differential equation for $\Psi$ with
periodic coefficients depending on $\Phi.$ For generic $\Phi$
it has only one solution, so $\Psi$ is determined by $\Phi$ up to
a constant $C.$ Reconstructing $\omega^{-1}_{k+1}$ 
from $\Psi,$ we obtain another
constant, so we have a following result: $\omega_{k+1}$ is obtained
from $\omega_{k}$ by a two-parametric correspondence. To
find $\omega_{k+1}$ starting from $\omega_k$, we must solve a first-order
linear differential equation to find $\omega^{-1}_{k+1},$ 
and then reconstruct $\omega_{k+1}$ from
$\omega^{-1}_{k+1}$ by inversion. 
The analogous situation was observed in the case
of $\mbox{SDiff}({\mathbb R}^2)$~\cite{MV2} which is integrable~\cite{FV}.
For this reason we consider system~(\ref{EL1},\ref{EL2})
to be a good candidate for an integrable Lagrangian discretization of the
Hunter-Saxton equation. See~\cite{P3} for more detailed discussion.

It should be remarked, that one can obtain the Hunter-Saxton equation
as an Euler equation not only on the Virasoro group, but also on
the group of orientation-preserving diffeomorphisms of the circle. This leads
to a particularly simple discretization 
$$
\left[\sqrt{(\omega_k)'}+%
\sqrt{(\omega^{-1}_{k+1})'}\right]'=0,
$$
which also has some nice properties, see~\cite{P3} for details.

\section*{Acknowledgments}

The author is very grateful to the Organizing Committee of the Workshop on
Superintegrability in Classical and Quantum Systems for giving him the
possibility to give this talk.

\end{document}